\documentclass[aps,prb,twocolumn,floats,epsfig,groupaddress]{revtex4-2}
\usepackage{amssymb}
\usepackage{amsbsy}
\usepackage{amsmath}
\usepackage{enumerate}
\usepackage{bm}
\usepackage{epsfig}
\usepackage{makecell}
\usepackage{color}
\usepackage{soul}
\usepackage{float}
\usepackage{braket}
\usepackage[colorlinks,linktocpage,bookmarks=false,citecolor=blue,linkcolor=red,urlcolor=blue]{hyperref}
\newcommand\mybar{\kern1pt\rule[-\dp\strutbox]{.8pt}{\baselineskip}\kern1pt}

\newcommand{\beq}{\begin{equation}}
\newcommand{\eeq}{\end{equation}}
\newcommand{\bea}{\begin{eqnarray}}
\newcommand{\eea}{\end{eqnarray}}

\begin{document}
\title{Relaxation of imbalance in a disordered XX model with on-site dephasing}
\author{Roopayan Ghosh}
\affiliation{Department of Physics and Astronomy, University College London,Gower Street, London}

\affiliation{Department of Physics, Faculty of Mathematics and Physics, University of Ljubljana, Jadranska 19, SI-1000 Ljubljana, Slovenia}

\author{Marko \v Znidari\v c}
\affiliation{Department of Physics, Faculty of Mathematics and Physics, University of Ljubljana, Jadranska 19, SI-1000 Ljubljana, Slovenia}
\begin{abstract}
  The relaxation of observables to their non equilibrium steady states in a disordered XX chain subjected to dephasing at every site has been intensely studied in recent years. We comprehensively analyze the relaxation of staggered magnetization, i.e., imbalance, in such a system, starting from the N\'eel initial state. We analytically predict emergence of several timescales in the system and extract results which match with large-system numerics without any extra fitting parameter until a universal timescale. An often reported stretched exponential decay is just one of the regimes which holds in a finite window of time and is therefore in fact not a true stretched exponential decay. Subsequently, the asymptotic decay of imbalance is governed by a power law irrespective of the disorder. We show that this emerges from the continuum limit of the low magnitude eigenspectrum of the Liouvillian. However, for finite systems, due to discreteness of the spectrum, the final phase of relaxation is governed by the relevant smallest Liouvillian gap.  
\end{abstract}
\maketitle
\section{Introduction}
Non interacting disordered systems in one dimension, isolated from external environment, are known to exhibit Anderson localization \cite{AndersonPhysRev.109.1492}. One expects that if a generic initial state is allowed to evolve in such a system, at long time scales, the wave function will not show significant change. Among the several quantifiers of this phenomenon \cite{MirlinRevModPhys.80.1355}, one of the most commonly used is imbalance, a quantity easy to measure in experiments. Imbalance $I$ is the staggered magnetization and captures the difference in orientation of spins on adjacent sites,
\begin{equation}
I=(1/L)\sum_{j=1}^L(-1)^j \langle\sigma_j^z\rangle,
\end{equation}
where $L$ is the length of the lattice, $\sigma^z$ is the Pauli spin z operator, and the expectation is taken with the state of which we want to measure the quantity. It should be evident that for computational states, the state with the largest $I$, which equals $1$, is the N\'eel state.

One can also surmise that for an Anderson localized system, if one starts from this state, one should see $I\sim 1$ at large timescales. However, if the localized system is no longer isolated, then the external degrees of freedom typically serve to break Anderson localization. The system eventually forgets the memory of its initial state. Accordingly, $I$ would also evolve with time. In the previous works, the focus has mainly been to find the non equilibrium steady state (NESS) in such systems \cite{Znidaric_2010,ScarPhysRevB.103.184202,PolettiPhysRevA.95.052107,PhysRevB.98.020202Flach,Wu_2019,Monthus_2017,PhysRevLett.118.070402Iva,Vershinina_2017}, or perturbations around NESS \cite{SchiroPhysRevB.104.144301}, while recently there has been some work studying how different observables relax to NESS in such systems \cite{MariyaMarkoPhysRevB.93.094205,GopalKnapPhysRevLett.119.046601,PolettiPhysRevLett.114.170401,PhysRevLett.116.237203Garrahan,10.21468/SciPostPhys.12.5.174BarLev,PhysRevB.95.024310Garrahan,PhysRevB.100.165144Poletti} or in Stark localized systems \cite{EckardtPhysRevLett.123.030602}. Of particular focus was the decay of imbalance. A slow stretched exponential ($A \exp(-t^{\alpha})$) decay of $I(t)$ has been reported \cite{AltmanPhysRevLett.116.160401,CaiBarthelPhysRevLett.111.150403,CaiPhysRevLett.124.130602,SchneiderPhysRevB.106.134211,PhysRevLett.116.237203Garrahan}, though the value of $\alpha$ obtained has some dispute. For example, a short theoretical analysis in Ref.~\onlinecite{CaiPhysRevLett.124.130602} finds $\alpha\sim 0.33$ whereas a different analysis in Ref.~\onlinecite{AltmanPhysRevLett.116.160401} put $\alpha\sim 0.5$. Numerical fits in Refs.~\onlinecite{AltmanPhysRevLett.116.160401} put $\alpha \sim 0.38$ and Refs.~\onlinecite{SchneiderPhysRevB.106.134211,PhysRevLett.116.237203Garrahan} put $\alpha \sim 0.42$.  Additionally, the analytical expressions suggested usually require at least one fitting parameter to be matched with the numerical results.

In our work, we seek to understand and resolve this inconsistency. Upon carefully analyzing the relaxation, we see that several time scales emerge. Broadly speaking one has a regime where off-diagonal matrix elements of time-dependent density matrix of the system, $\rho(t)$ are large, then a regime in which $\rho(t)$ becomes increasingly diagonal due to dephasing and the scaling variable is~\cite{MariyaMarkoPhysRevB.93.094205} $\tau=8\gamma t/W^2$ ( where $\gamma$ and $W$ denote the strength of dephasing and disorder respectively), and finally a regime where the system starts to ``feel" its finite size $L$. In fact, we argue that the ``stretched exponential" is not a true (asymptotic) description for the relaxation. It holds only in a finite window and is universal and independent of disorder type just in a linear-expansion regime, after which a non generic disorder-dependent behavior follows. Additionally, beyond a timescale, $\tau=8 \gamma t/W^2\sim 5$, we show that the relaxation first smoothly changes to $1/t^{3/2}$ irrespective of boundary conditions. Subsequently, only for open boundary conditions, at large $\tau \sim L$, the decay asymptotically changes to $1/(L\sqrt{t})$, which is also the standard result for clean open boundary systems with dephasing. Finally, for finite size systems, an exponential decay occurs at the longest timescale, governed by the finite Liouvillian gap.

Using a toy model involving three lattice sites, we find an expression of $I(\tau)$ analytically, using second order perturbation theory, that matches with numerics very accurately, until $\tau\sim 1$, without any fitting parameters. From the model, we show that the universal behavior $I(\tau)=(1-\beta \sqrt{\tau})$ occurs for very short times $\tau\ll 1$, beyond which the non universal behavior follows until $\tau\sim 1$. 

We first provide a brief description of our main results in Sec.~\ref{summary} before going into the details later.

\section{Summary of Results}
\label{summary}
\begin{figure}
\centering
\includegraphics[width=0.85 \columnwidth]{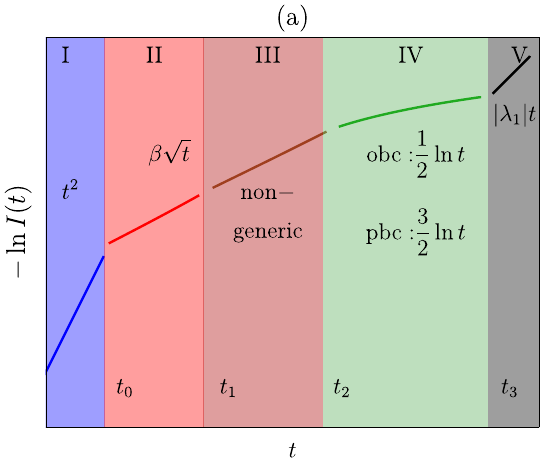}
 \includegraphics[width=0.9 \columnwidth]{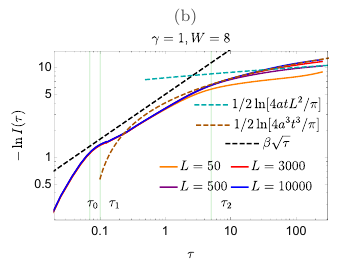}
 \caption{(a) Schematic figure denoting the five regimes of imbalance decay for an initial N\'eel state. Panel (a) shows the schematic diagram with timescales $t_0\sim {\rm Min}(1/\gamma,1/W)$, $t_1\approx 0.1 W^2/(8 \gamma J^2 )$, $t_2 \approx  5 W^2/(8 \gamma J^2 )$ and $t_3 \sim L^2 W^2/\gamma$, whereas in (b) disorder averaged exact numerical data (open boundary conditions) for different system sizes $L$ are plotted as a function of $\tau=8 \gamma t/W^2$ ($J=1$, $\gamma=1$, $W=8$). Black dashed line is a stretched exponential with $\beta$ obtained from expanding Equation~\eqref{box}, while brown and green dashed curves are the theoretical predictions of Equation~\eqref{pbclaw} and Equation~\eqref{obclaw} ($L=500$). The green vertical lines denote the time stamps of the schematic diagram for $L=500$ ($\tau_3$ falls outside the plot's range).
 }
\label{fig1}
\end{figure}

We begin by summarizing our findings. Starting from the N\'eel initial state, in the XX model with on-site disorder and dephasing, the behavior of $-\ln I(t)$ can be schematically represented by Fig.~\ref{fig1}(a), where $\ln$ denotes the natural logarithm, $\gamma$ is the strength of dephasing, $W$ is the strength of disorder and $J$ is the exchange interaction strength of XX model.

As evident from the figure, five windows with different behavior emerge during the evolution of $I(t)$. Depending on $t$ (or the scaled $\tau=8\gamma t/W^2$) they are as follows.
\begin{enumerate}[(I)]
\item $t < t_0 \sim {\rm Min} (1/\gamma,1/W)$: the shortest timescale in the problem. Here, $ I(t) \sim 1-t^2$ and hence $-\ln I(t) \sim t^2$. During this time period, off diagonal correlations first develop in the system and then start decaying after reaching a maxima.\footnote{Recall that since we start from the N\'eel state, a computational state, hence at $t=0$ the system is completely uncorrelated.} When the disorder strength $W$ is small compared to the dephasing $\gamma$, $t_0$ is given by $\sim 1/\gamma$, akin to clean systems. However when $W \gg \gamma$, then this behavior exists until $t \sim 1/W$. 
\item $t_0 < t < t_1 \sim 0.1 W^2/(8 \gamma J^2 )$: this is where local relaxation of $\sigma^z$ starts, and most (but not all) off-diagonal correlations become negligible. In this timescale one sees $-{\rm \ln} I(t)\sim -{\rm \ln}(1-\beta \sqrt{t}) \sim \beta \sqrt{t}$ behavior irrespective of the nature of disorder chosen ($\beta$ can depend on nature of the disorder). This is the short time, linear regime of the stretched exponential behavior noted in previous works\cite{AltmanPhysRevLett.116.160401,CaiBarthelPhysRevLett.111.150403,CaiPhysRevLett.124.130602,SchneiderPhysRevB.106.134211,PhysRevLett.116.237203Garrahan}. Unlike previous results, we find that this regime is not asymptotic and is only valid until $\tau\sim\tau_1=(8 \gamma J^2 t_1)/W^2 = 0.1$. Therefore, the ``stretched exponential" is not really a true stretched exponential decay as it does not hold asymptotically, i.e., at arbitrarily small values of $I$.
\item  $ t_1<t<t_2\approx 5 W^2/( 8 \gamma J^2)$: in this regime, which holds until $\tau\sim\tau_2= (8 \gamma J^2 t_2)/W^2  \approx 5$ a non generic decay dependent on the choice of disorder distribution is seen. $t_2$ also marks the end of local relaxation in the system.
 
  \item $t_2<t< t_3$: in this regime the decay is a power law due to a continuum of low magnitude eigenvalues of Liouvillian \cite{CaiBarthelPhysRevLett.111.150403,MedvKehrinPhysRevB.90.205410}. Because details of eigenvalues and overlaps depend on boundary conditions, i.e., having periodic or open boundary conditions, the imbalance decay can likewise depend on it. Specifically, for periodic boundary conditions one has $L$-independent $I(t)\propto 1/t^{3/2}$, while for open boundary conditions one eventually gets an $L$-dependent decay $I(t)\propto 1/(L\sqrt{t})$. Because the imbalance for open and periodic boundary conditions agree up to extensive times $t \sim L$ (light-cone hitting the boundary), one has an interesting situation for open boundary conditions: for sufficiently large $L$ one will initially have $I(t) \sim 1/t^{3/2}$, which will then gradually transition into the asymptotic decay $I(t)\propto 1/(L\sqrt{t})$ at a time $\tau \approx L/2$ [in Fig.~\ref{fig1}(b) this time is $\tau \approx 25$ for $L=50$ and $250$ for $L=500$].
    
\item $t> t_3 \sim L^2W^2/\gamma$: the final timescale in finite sized systems is governed by the Liouvillian gap, i.e., the largest real nonzero eigenvalue of the Liouvillian (since it has an eigenvalue of zero). Here the decay is given by $I(t)\sim \exp(-|\lambda_1| t)$ where $\lambda_1$ is the Liouvilian gap. This timescale begins in the leading order around $t_3 \sim 1/|\lambda_1|$. As will be discussed in Sec.~\ref{region5}, $\lambda_1\sim \gamma/(L^2W^2)$ (for open and periodic boundary conditions); hence this regime starts around $t_3 \sim L^2 W^2/\gamma$.
\end{enumerate}

It is worth noting that until Region IV there is no system size dependence of the results; neither the time windows nor the behavior depend on $L$. However, the ending time $t_3$ of region IV and the behavior of $I(t)$ for open boundaries depend on $L$ (for periodic boundary conditions $I(t)$ does not depend on $L$). This feature is clearly visible in Fig.~\ref{fig1}(b) where for $\tau>5$ different system sizes start showing different behavior. We perform disorder averaging over $10^2$ realizations for $L=50,500$ and $10$ for $L=3000$. 

Since until $\tau\sim 5$, the rescaled data for the different system sizes clearly overlap with one another and are almost linear in log-log scale, this prompted the stretched exponential fits in previous works. In Fig.~\ref{fig1}(b), we have added a black dashed line showing the stretched exponential expression obtained from a theoretical computation in Sec.~\ref{region23}. As will be clear from the analysis in that section, this behavior is expected to hold best until small $\tau \sim 0.1$, which is what is demonstrated in the plot. \footnote{Hence numerical fits of the form $\exp(-t^{\alpha})$ in Refs. \onlinecite{AltmanPhysRevLett.116.160401,CaiBarthelPhysRevLett.111.150403,CaiPhysRevLett.124.130602,SchneiderPhysRevB.106.134211} found different $\alpha$'s depending on fit windows as the behavior is not a true stretched exponential. } Finally for $L=50$ one sees a fast growth of $-\ln I(t)$ (see a slight bend upwards) at $\tau>200$, which is due to the exponential decay of $I(t)$ from the finite size Liouvillian gap.

In what follows, we shall discuss the above findings in detail and provide a theoretical understanding for the same. In the next section, we describe the model in more detail and elaborate on the technique used to compute the numerical results. Then in Sec.~\ref{timescales} we elaborate on the behavior of $I(t)$ in different timescales. Finally, in Sec.~\ref{discussion} we provide some final remarks on our results and possible extensions of this work.
\section{Model}
\label{model}
We take the one dimensional disordered XX chain with spin 1/2 particles,
\begin{equation}
H=-J\sum_{j=1}^{L-1}(\sigma^x_j\sigma^x_{j+1}+\sigma^y_j\sigma^y_{j+1})+\sum_{j=1}^L h_j\sigma^z_j 
\label{lind1}
\end{equation}
where $h_j$ are the on-site disorders and $\sigma^{x,y,z}$ denotes the Pauli matrices. In this work, unless otherwise mentioned, $h_j$s are chosen from a uniform distribution in $(-W,W)$, with $W$ being the strength of disorder. We set $J=1$ for the rest of our work. To simulate an open system, each spin is exposed to a dephasing. The evolution of the system's density matrix is given by
\begin{equation}
\frac{{\rm d}\rho}{{\rm d} t}=i [\rho,H]+\mathcal{L}^{{\rm deph}}(\rho)=\mathcal{L}(\rho)
\end{equation}
The nonunitary part can be written in terms of Lindblad operators as, $\mathcal{L}^{{\rm deph}}(\rho)=\sum_k([L_k \rho,L_k^{\dagger}]+[L_k, \rho L_k^{\dagger}])$. We choose $L_k=\sqrt{\frac{\gamma_k}{2}}\sigma_k^z$ to represent the on site dephasing term. For the rest of this work we shall put $\gamma_k=\gamma$, i.e. have a spatially uniform dephasing. This choice of dephasing causes an exponential decay of the off diagonal elements of the density matrix with a strength $\gamma$, in the diagonal basis of $\sigma^z$ in the absence of disorder.

If we want to solve Equation~\eqref{lind1} directly, numerically or otherwise, we need to solve a set of $4^L-1$ coupled differential equations. Fortunately due to the choice of dephasing and the quadratic nature of the Hamiltonian, the exponentially many equations can be decoupled into blocks of polynomial complexity. \cite{Znidaric2013,Znidaric_2010,PolettiPhysRevA.98.052126,Eisler_2011,Temme_2012,CiracPhysRevA.87.012108,arxiv.2210.10856,SchiroPhysRevB.104.144301} In other words, observables follow a hierarchy based on the number of fermionic operators they contain. For example, the block of two point correlators decouples from the rest, and one can write a closed set of equations for these observables. Then this solution serves as a source term for the three point correlations and so on.  Since we are interested in imbalance, which can be extracted from a two-point correlator in the fermionic language $(\sigma^z \sim c^{\dagger} c)$, we will just consider the subspace of two point correlation functions. Following the consideration of Refs.~\onlinecite{Znidaric_2010,Znidaric2013}, we define the operators,
\begin{eqnarray}
A(t)&=&\sum_{r=1}^L \sum_{j=1}^{L+1-r}a_j^{(r)}(t)A_j^{(r)}\nonumber \\
B(t)&=&\sum_{r=2}^L \sum_{j=1}^{L+1-r}b_j^{(r)}(t)B_j^{(r)},
\label{hierarchy}
\end{eqnarray}
where $A_j^{(r+1)}=\sigma_x^j Z_{j+1}^{(r-1)} \sigma_x^{j+r}+\sigma_y^j Z_{j+1}^{(r-1)} \sigma_y^{j+r}$ and $B_j^{(r+1)}=\sigma_x^j Z_{j+1}^{(r-1)} \sigma_y^{j+r}-\sigma_y^j Z_{j+1}^{(r-1)} \sigma_x^{j+r}$ for $r\ge 2$. $Z_j^{(r)}=\sigma_z^j\hdots \sigma_z^{j+r-1}$ are strings of $\sigma_z$ operators, and $A_j^{(1)}=-\sigma_z^j$. Then the equation governing the time evolution  of the set of two point correlation functions can be written compactly as 
\begin{equation}
\frac{{\rm d}\mathbf{C}(t)}{{\rm d} t}+2 {\rm i} (\mathbf{P}\mathbf{C}(t)-\mathbf{C}(t)\mathbf{P}^T)+2(\mathbf{\Gamma}\tilde{\mathbf{C}}(t)+\tilde{\mathbf{C}}(t)\mathbf{\Gamma})=0, 
\label{diffeq}
\end{equation}
where $\mathbf{C},\tilde{\mathbf{C}},\mathbf{P},\mathbf{\Gamma}$ are $L \times L$ matrices. Their elements are defined as, $C_{jk}(t)=a_j^{(k-j+1)}(t)+{\rm i} b_j^{(k-j+1)}(t)$ for $k>j$, $C_{jj}(t)=a_j^{(1)}(t)$ and $C_{jk}=C_{kj}^*$. $\tilde{\mathbf{C}}=\mathbf{C}-{\rm diag} (\mathbf{C})$. $\mathbf{P}=\mathbf{W}-\mathbf{T}$ where, $W_{jk}=h_k \delta_{jk}$, the on-site disorders and for our model. Also for our model, $T_{jk}=J(\delta_{j,k-1}+\delta_{j,k+1})$, as we consider only nearest neighbor couplings \footnote{For fermionic Hamiltonians, if longer range couplings are present then, in the most general case,  $T_{jk}=J_{jk}$ where $J_{jk}$ denotes the hopping strength between two different sites at positions $j$ and $k$.} and  $\Gamma_j^k=\gamma \delta_{jk}$. Clearly, $I(t)=(1/L)\sum_j(-1)^{j+1}C_{jj}(t)$. Hence for the N\'eel initial state $C_{jj}(0)=(-1)^{j+1} $ and $I(0)=1$.

Equation~\eqref{diffeq} can be recast into a linear differential equation with $L^2$ variables, in the form
\begin{equation}
\frac{d\mathbf{f}}{dt}=\mathcal{Q} \mathbf{f}
\label{linearised}
\end{equation}
where $\mathbf{f}=(C_{11},C_{12},\hdots,C_{1L},C_{21},\hdots,C_{LL})$ and $\mathcal{Q}$ is the $L^2 \times L^2$ matrix governing the evolution. In fact, due to the hierarchical structure of observables, the eigenvalues of $\mathcal{Q}$ are exactly the eigenvalues of the one particle sector of the Liouvillian, $\mathcal{L}$, while the eigenvectors of both are connected via an appropriate rotation. This linearized equation will be useful in understanding the behavior of $I(t)$ in different regimes.

Unlike most of the previous results for this model which are generated by either an effective Hamiltonian or DMRG based techniques \cite{AltmanPhysRevLett.116.160401,CaiPhysRevLett.124.130602}, we use Equation~\eqref{diffeq} or Equation~\eqref{linearised} for the analysis that follows. This not only allows us to obtain an exact description of the correlators but also provides access to sizes and times an order of magnitude larger than usually accessed by DMRG techniques. Thus we can make better statements about system size dependence of the results.

\section{Different timescales}
\label{timescales}

Let us now discuss each timescale in more details.
\subsection{Region I}
\label{reg1}
\begin{figure}
\includegraphics[width=0.75 \columnwidth]{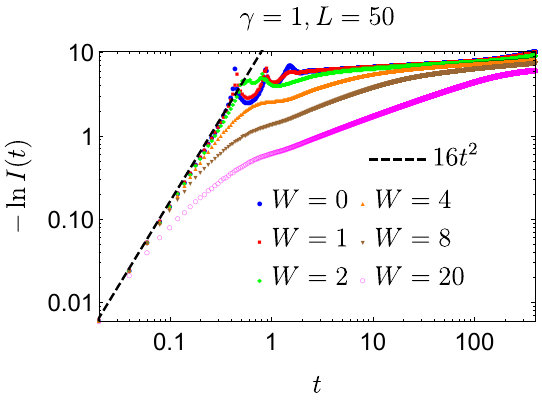}
\caption{Initial $t^2$ behavior of $-\ln I(t)$ in region I for different disorder strengths $W$ and open boundary conditions (obc).}
\label{fig2}
\end{figure}
Doing a simple power series expansion of $I(t)$ in $t$, the first nonzero term turns out to be quadratic. Hence, we expect $I(t)\sim 1-\eta t^2$ behavior at the smallest timescale. As evident from Fig.~\ref{fig2}, this is the case. This behavior continues until time $t_0$ which is dependent on $\gamma$ and $W$. To be more precise, let us first consider the simpler case, $W\rightarrow 0$. At $t=0$, since we start from a pure state, $\rho(t=0)$ has only one nonzero element, which is in the diagonal, and $C(t=0)$ is diagonal as well. Correlations then spread rapidly throughout the system, reach a maxima and start decaying around $t \sim 1/\sqrt{|\gamma^2-4|} \sim 1/\gamma$ for $\gamma \gg 1$. As we see from Fig.~\ref{fig2}, indeed below $t_0 \sim 0.5$ for $\gamma=1$ and $W=0$, one can approximate $I(t)$ as $1-\eta t^2$  and hence $-\ln I(t)\sim \eta t^2$. From the numerical fit in Fig.~\ref{fig2} shown by the black dashed line, we find $\eta \sim 16$. This behavior can also be qualitatively extracted from a simple two-site model (details in Appendix~\ref{appA1}) and we get, for small $t$, $ I(t) \sim 1-8 t^2 +O(t^3)$. One can then numerically check that on addition of a few more sites to the system  $\eta$ approaches $16$ rapidly.

However, on addition of disorder, $t_0$ reduces with increasing $W$, as can be seen Fig.~\ref{fig2}. This can be qualitatively understood from Equation~\eqref{diffeq}, where we see that correlations can develop in the system either due to $\mathbf{P}$, i.e. the terms involving disorder or due to $\mathbf{\Gamma}$ which involves dephasing. The dominant term would then determine the smallest timescale. Also, it can be shown using the two-state model again that, for $W \gg \gamma$, the $t_0$ now becomes $1/\delta$, with $\delta$ being the difference of on-site disorders in the two-site problem.  Hence several timescales emerge due to different $\delta$'s at different lattice sites and we will get $t_0 \ll 1/\gamma$. If we approximate the width of the distribution of $\delta$s as $W$, we can say that $t_0$ is around $ {\rm Min}(1/\gamma,1/W)$. This is region I.

\begin{figure*}
\includegraphics[width=0.9 \columnwidth]{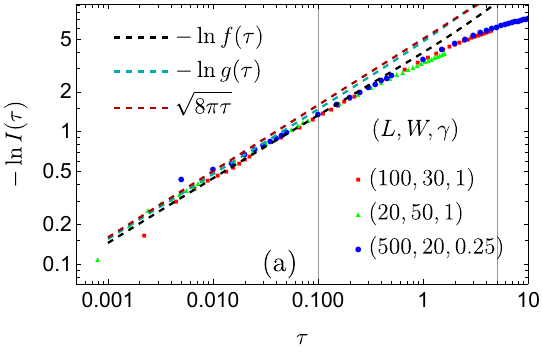}
\includegraphics[width=0.9 \columnwidth]{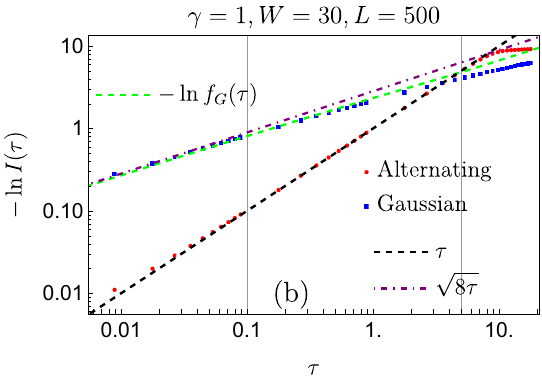}
\caption{ Comparison of analytical predictions with exact numerics in regions II and III for obc. (a) Plot showing how the rescaled time causes complete collapse of data with different $L,\gamma, W$, and agreement of the said numerical results for the box distributed random on-site disorder, with the theoretical predictions. $f(\tau)$ is defined in Equation~\eqref{firstapprox} and $g(\tau)$ is defined in Equation~\eqref{box}. (b) Plot showing agreement of our theoretical prediction with numerical results derived from for two other disorder  distributions, viz.  Gaussian random and alternating disorder, see Equation~\eqref{gaussian} and Equation~\eqref{alternating} respectively.  $f_G(\tau)$ is obtained from Equation~\eqref{firstapprox} where the random numbers are picked from a Gaussian distribution. Vertical lines show the starting and ending times of region III.}
\label{fig3}
\end{figure*}
\subsection{Regions II and III}
\label{region23}
Next, we shall understand the decay of imbalance in regions II and III. Local evolution is dominant in these regimes and we can use the same analysis to describe both. Hence we discuss them together. A timescale separating the two regimes will emerge naturally from the computation that follows. 

We consider $W \gg \gamma$, a regime where localization effects would be prominent in the system. Then, to understand the evolution of $\langle\sigma_j^z \rangle$ for site $j$, we need to take into account the influence of the neighboring sites, $j-1$ and $j+1$. Matrix $\mathbf{C}$ in Equation~\eqref{hierarchy} for this three-site system, becomes a $3 \times 3$ matrix. We further ignore $C_{ij}$'s where $|j-i|>1$, as they are negligible compared to the rest in this timescale. We can then use Equation~\eqref{linearised} with
\begin{equation}
 \mathbf{f}=\begin{pmatrix}
 C_{11}, & C_{12}, & C_{21}, & C_{22}, & C_{23}, & C_{32}, & C_{33}
 \end{pmatrix},
 \label{threesitef}
 \end{equation}
 where we have taken $j=2$ without loss of generality. The form of $\mathcal{Q}$ for this system in Equation~\eqref{linearised} is given in Appendix~\ref{appA2}. We use second order perturbation theory to find the eigenvalues as outlined in Appendix~\ref{appA2}, since exact analytical diagonalization is not tractable. We also just need to diagonalize the subspace where we have the smaller modulus eigenvalues, as the larger modulus eigenvalues dictate the behavior in regime I. Doing so, we obtain the relevant eigenvalues as given in Equation~\eqref{eigenvalues} in Appendix~\ref{appA2}. These in the regime $ W \gg \gamma$ can be approximated as $\Delta^{(0)}=0$ and
\begin{equation}
\Delta^{(\pm)}\sim\frac{8 \gamma  (\delta_1^2+\delta_2^2\pm\sqrt{\delta_1^4+\delta_2^4-\delta_1^2\delta_2^2})}{\delta_1^2 \delta_2^2},
\label{approxeigen}
\end{equation}
where $\delta_1=|h_{1}-h_2|$ and $\delta_2=|h_2-h_{3}|$. 
Hence the long time behavior of $C_{22}(t)=-\langle\sigma_2^z(t)\rangle$ can be generically written as, 
\begin{equation}
C_{22}(t)=\sum_{m=\pm,0}d^{(m)}_2 \exp(-\Delta^{(m)} t)
\label{C22three}
\end{equation} 
where $d^{(m)}_2$ is the corresponding element of the $m^{th}$ eigenvector, weighed by the factors arising from  the initial conditions. We shall drop the subscript $2$ in what follows. 

Further simplification of Equation~\eqref{C22three} can be made by observing that typically (see Appendix~\ref{appA2}) $d^{(+)}$ is the largest coefficient, hence the principal mode of relaxation is $\Delta^{(+)}$. The contribution from $\Delta^{(0)}$ and $\Delta^{(-)}$ modes weighted by $d^{(0)}$ and $d^{(-)}$, respectively, effectively act as perturbations around the principal mode\footnote{This is difficult to rigorously prove analytically, but can be seen numerically} (see Appendix~\ref{appA2}). In the random disorder case since $\Delta^{(+)}$'s are different for different sites, we reintroduce the subscript $j$, and write for the N\'eel initial state, $ I(t)=(1/L)\sum_{j=1}^L|C_j(t)| \sim(1/L)\sum_{j=1}^L |d_j^{(+)}| e^{-\Delta^{(+)}_j t}$, where the perturbative corrections due to the other modes can be neglected due to averaging.
 Additionally, since $d_j^{(+)}$'s are $\mathcal{O}(1)$(see Appendix~\ref{appA2}), it is not essential to keep track of them. Hence, given a disorder distribution one can calculate the evolution of imbalance for the N\'eel initial state in the thermodynamic limit as
 \begin{equation}
 I(t)= f(t,W,\gamma)=\lim_{L \to \infty}(1/L)\sum_{j=1}^L e^{-\Delta^{(+)}_j t},
 \label{firstapprox}
\end{equation}
where different $\Delta^{(+)}_j$s are obtained from the distribution of $h_j$s, using Equation~\eqref{approxeigen}. From Equation~\eqref{firstapprox} we can see that in this timescale, $I(t)$ is the average over the decay of magnetization at individual sites with effective decay rates at each site controlled by the disorders in its nearest neighbors. From Equation~\eqref{approxeigen}, we also see an emergent energy scale $8 \gamma/W^2$ (recall $J=1$), which gives us the rescaled time $\tau=8 \gamma t/W^2$.

  In Fig.~\ref{fig3}(a) we show comparison between exact numerical results with Equation~\eqref{firstapprox} and other approximations described later in the section. The plots for different system sizes $L$, disorder strengths $W$, and dephasing $\gamma$ obtained via exact numerics collapse on each other when we use rescaled $\tau=8 \gamma t/W^2$, confirming the presence of the universal timescale\cite{MariyaMarkoPhysRevB.93.094205}. As shown in Fig.~\ref{fig3}(a) with the black dashed line, computing Equation~\eqref{firstapprox} by sampling a large number of random numbers from the box disorder distribution, we get a result that is a very accurate match with exact numerics until $\tau \sim 1$. Note how a numerical result obtained via solving many coupled differential equations can be replicated by correctly sampling the underlying disorder distribution, due to locality of the evolution.
 
However, it is difficult to perform the summation analytically to obtain a closed form expression for different disorder distributions. To proceed further we need to go to the continuum limit and write Equation~\eqref{firstapprox} as $I(t)=\int d \Delta^{(+)} p(\Delta^{(+)}) e^{-\Delta^{(+)}t}$, with $ p(\Delta^{(+)})$ being the probability distribution of $\Delta^{(+)}$. Even so, computing $p(\Delta^{(+)})$ analytically is a very difficult task. Hence, we need to make further simplifying assumptions. If we approximate $ \delta_1^4+\delta_2^2-\delta_1^2\delta_2^2\sim(\delta_1^2+\delta_2^2)^2$ and hence $\Delta^{(+)} \sim 16\gamma ( \frac{1}{\delta_1^2}+\frac{1}{\delta_2^2})$, the integral becomes tractable. This approximation is valid in the limit of $|\delta_1^2- \delta_2^2| \gg 0$, where we have $\delta_1^2\delta_2^2 \ll \delta_1^4+\delta_2^4$ and hence works in the large $\Delta^{(+)}$ tail of $p(\Delta^{(+)})$ very well; see Fig.~\ref{fig6a} in Appendix.~\ref{appA2}. \footnote{This choice of approximation is used to obtain a tractable form for $p(\Delta^{(+)})$. Other choices are also possible in the said limit, but they do not simplify the calculation. We show a comparison of the probability distribution of the approximated and exact $\Delta^{(+)}$ in Appendix~\ref{appA2}}. Then, we consider that $\delta_1$ and $\delta_2$ are independently drawn from different sets of random numbers to avoid calculating the complicated convolution term. This allows us to write
\begin{equation}
I(t)=\left(\int d\delta p(\delta) e^{-16 \gamma t/\delta^2}\right)^2
\label{master1}
\end{equation}
where $p(\delta)$ is the probability distribution of $\delta$s. Equation~\eqref{master1} allows us to compute the evolution of $I(t)$ analytically for any given disorder distribution. For example, for a disorder distributed uniformly in $(-W,W)$, i.e. the box distribution, Equation~\eqref{master1} gives the result in terms of known mathematical functions as,
\begin{eqnarray}
I(\tau=8 \gamma t/W^2)=g(\tau)&=&[-\sqrt{2 \pi } \sqrt{\tau } \text{erfc}\left(\sqrt{\frac{\tau }{2}}\right)\nonumber \\
&&+e^{-\frac{\tau }{2}}+\frac{1}{2} \tau  \Gamma \left(0,\frac{\tau }{2}\right)]^2
\label{box}
\end{eqnarray}
where $\text{erfc(z)}=1-\text{erf}(z)$, ${\rm erf}$ denotes the Error function and $\Gamma$ is the incomplete gamma function defined by $\Gamma(0,z)=\int_z^{\infty}e^{- t}/t dt$. The green dashed line in Fig.~\ref{fig3}(a), obtained from Equation~\eqref{box} shows good agreement until $\tau \sim 0.1$, which is expected since our approximation works best in the large $\Delta^{(+)}$ tail.

\paragraph*{\textbf{$1-\beta \sqrt{t}$ scaling regime:}} Actually, in the regime of $\tau \ll 1$ or $t \ll W^2$, Taylor expanding Equation~\eqref{box} in $\tau$, we get $I(\tau)=1-\beta\sqrt{\tau} +O(\tau)$ with $\beta=\sqrt{8\pi}$, which is exactly the linear term in the stretched exponential\cite{CaiBarthelPhysRevLett.111.150403}  $e^{-\beta \sqrt{\tau}}$. In Fig.~\ref{fig3}(a), we plot $-\ln I(\tau)\sim \beta \sqrt{\tau}$ as the brown dashed line, and it also agrees well with exact numerics until $\tau \sim 0.1$. 

Generally, if $p(\delta)$ is an analytic function of $\delta$,  then it can be expanded as $p(\delta)=\frac{1}{\mathcal{N}}(c+O(\delta))$, where $\mathcal{N}=\int d\delta (c+O(\delta)) $. Since $\int d\delta e^{-16 t/\delta^2}\sim t^{1/2}+O(t)$, the lowest order term obtained from $p(\delta)$ in this form would be $\sqrt{t}$ (for $c=0$ the result would be different).  For many different disorder distributions this is a good approximation of results until higher order terms become important with larger $t$. 

To highlight this special scaling, we label the regime where the linear approximation of $\beta \sqrt{\tau}$ is valid as region II, which is approximately until $\tau=0.1$, and denote the nongeneric regime during $0.1\lesssim\tau \lesssim 5$ as region III.
 
\paragraph*{\textbf{Other disorder distributions:}} To further consolidate our claims, we repeat the above calculations for other disorder distributions, viz. the Gaussian random distribution and staggered on-site or alternating potential, shown in Fig.~\ref{fig3}(b). We have also checked the quasiperiodic Aubre Andre distribution; the data are not shown to avoid cluttering. In Fig.~\ref{fig3}(b), results from Equation~\eqref{firstapprox} are represented by the green dashed line for the Gaussian disorder case and the black dashed line for the alternating potential case. In both cases (and the quasiperiodic one not shown), we see that Equation~\eqref{firstapprox}, obtained from the second order perturbation result of the three-site model, captures the nuances in the evolution very well until $\tau=1$ (longer for the alternating potential see Appendix~\ref{appA2}). In general, our analysis is valid for any disorder distribution where typical $|h_j-h_{j+1}| \gg0$ to make the second order perturbation theory hold.\footnote{One can still numerically solve the small size Hamiltonians to obtain the eigenenergies nonperturbatively and continue with the analysis.} 

For the Gaussian distribution it is also possible to compute Equation~\eqref{master1} analytically and the result is
\begin{eqnarray}
I(\tau)=&e^{- \sqrt{8\tau}},
\label{gaussian}
\end{eqnarray}
which is a different result to Equation~\eqref{box} and exactly the stretched exponential\footnote{Reference~\onlinecite{AltmanPhysRevLett.116.160401} arrived at this result using a different approach effectively leading to the same integral. However, under their approximations one arrives at this result irrespective of the disorder chosen and then needs to fix a free parameter via numerical fitting to fit to different disorders}. This is shown as the purple dot-dashed line in Fig.~\ref{fig3}(b), which we again see to be valid until $\tau\sim 0.1$, thus showing a separation of timescales similar to box distribution. Other standard disorder distributions such as the exponential and student's T distribution also show similar timescale separation.

Finally, for the staggered potential, i.e., when $h_j=(-1)^j W$, Equation~\eqref{firstapprox} has a simple form. Since we have $\delta_1=-\delta_2=\delta/2= W$, $\Delta^{(+)}=6 \gamma/W^2$ for all sites. In this case the weight of the $\Delta^{(-)}$ mode $d^{(-)}=0$; hence $I(t)\sim d^{(0)}+d^{(+)} e^{-\Delta^{(+)} t}$. For large $W$, this can be approximated as $f_{{\rm alt}}=e^{-\Delta_{{\rm eff}} t}$, where  $\Delta_{{\rm eff}}=\frac{8 \gamma}{W^2}$ and hence 
\begin{equation}
I(\tau) \sim e^{-\tau}.
\label{alternating}
\end{equation} See Appendix~\ref{appA2} for more details. In Fig.~\ref{fig3}(b), we see exactly the expected behavior. The almost perfect agreement between our theoretical predictions with the numerical data provides a good benchmark about the generality of our theory.
 
\paragraph*{\textbf{Effective Hamiltonian approach:}} Before we end this section we make a digression to briefly discuss an alternative approach.  In Ref. \onlinecite{MariyaMarkoPhysRevB.93.094205} it was shown that, for the interacting XX chain with dephasing and strong disorder, beyond a timescale given by $1/\gamma$, off-diagonal terms of the density matrix are negligible and the evolution of the diagonal terms of the density matrix is given by the differential equation, $\frac{d \bm{\rho}_D}{dt}=-\mathcal{H}_{\rm eff}\bm{\rho}_D$, where $\bm{\rho}_D$ denotes the diagonal part of $\bm{\rho}$ and 
\begin{equation}
\mathcal{H}_{\rm eff}=\sum_{j=1}^L 2 \frac{(\gamma_j+\gamma_{j+1})}{(h_j-h_{j+1})^2+(\gamma_j+\gamma_{j+1})^2}(1-\bm{\sigma}_j.\bm{\sigma}_{j+1}),
\label{effham}
\end{equation}
 where $\bm{\sigma}=(\sigma^x,\sigma^y,\sigma^z)$. An emergent rescaled time, $ 4\gamma t/(W^2+4 \gamma^2)$ is seen as the overall prefactor of $\mathcal{H}_{\rm eff}$, if we consider $\gamma_{j}=\gamma$ and $\delta_j=h_j-h_{j+1}\sim W$, reminiscent of the one we extracted from the three site model, since this is also a second order perturbative description. In our non-interacting XX model, this effective description still holds as there is no term involving any interaction in Equation~\eqref{effham}. Beyond a cutoff timescale, which for $L=24$ and $W=30$ is given by $\tau \sim 0.01$ (check Appendix~\ref{appB}) the agreement of exact numerics and evolution of $I(t)$ with $\mathcal{H}_{\rm eff}$ is excellent. This means the effective Hamiltonian describes the system approximately for $\tau \gg \tau_0$ in our model ($\tau_0 \sim 0.001$ for $W=30$), while before that timescale the off diagonal elements of the density matrix still have significant contribution to the evolution. While, a direct use of $\mathcal{H}_{\rm eff}$ to repeat the analysis of this section to understand regions II and III is a bit involved, we shall use this effective Hamiltonian to explain the behavior in region IV and V in the following sections.

Finally, it is worth noting that for $W=0$, regions II and III no longer exist and there is an oscillatory behavior in $-{\rm \ln} I(t)$ as can be seen in Fig.~\ref{fig2}, with an envelope growing as $4 \gamma t$. For large $\gamma$ one can again apply the effective Heisenberg model to obtain these results for this system as in Ref.~\onlinecite{CaiBarthelPhysRevLett.111.150403}. Upon analysis via our three-site model, we observe that the magnitude of the real part of eigenvalues involved in the evolution $\sim 4\gamma$ as the corresponding eigenvectors carry almost all the weight--- a distinct shift from what happens in the disordered case, where lower magnitude eigenvalues carry most of the weight. Near the end of this timescale the system begins to realize its nonlocal nature and the evolution slowly changes to what is seen in region IV, discussed below, irrespective of the presence of disorder.

\subsection{Region IV}
\label{region4-1}
\begin{figure*}
\includegraphics[width=0.9 \columnwidth]{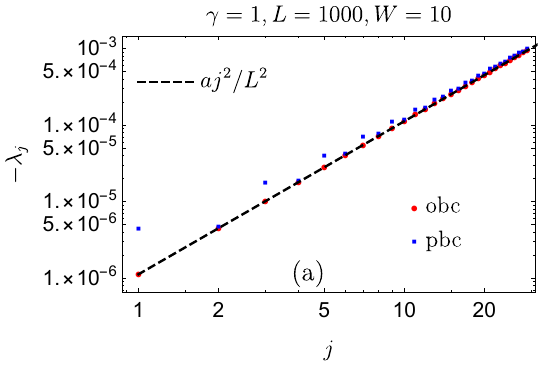}
\includegraphics[width=0.9 \columnwidth]{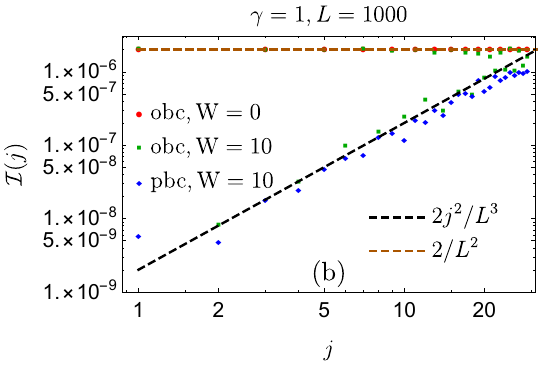}
\includegraphics[width=0.9 \columnwidth]{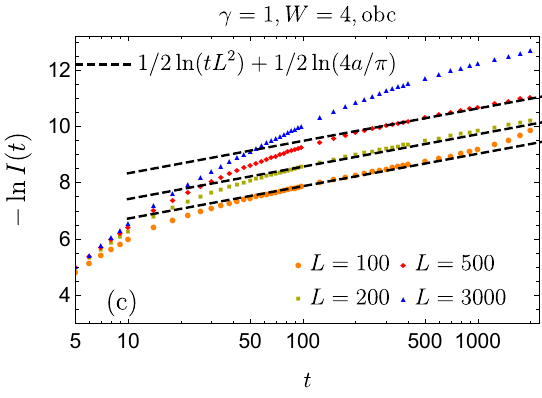}
\includegraphics[width=0.9 \columnwidth]{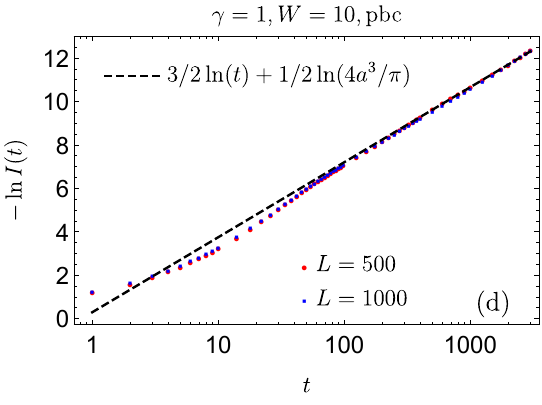}
\caption{Power law decay of $I(t)$ in region IV for obc systems. (a) Plot showing $j^2$ scaling of low magnitude Liouvillian eigenvalues averaged over $100$ disorder realizations compared with approximate theoretical prediction in Equation~\eqref{eigenvaluesapp}. (b) Plot showing scaling of $\mathcal{I}(j)$ defined in Equation~\eqref{support} averaged over $100$ disorder realizations. obc and pbc denote open and periodic boundary conditions respectively. See text for details. (c) Plots showing the scaling of $-\ln I(t)$ with $t$ for obc at different system sizes $L$. (d) Same as (c) but for periodic boundary conditions.}
\label{fig4}
\end{figure*}

\begin{table*}
\centering
\begin{tabular}{|c|c|c|c|c|c|}
\hline
Case & disorder & $\alpha$ & $\beta$& $I(t)$ &$-\ln I(t)$\\
\hline
pbc, even $L$ &$0$& $2$ &no overlap & $e^{-4 \gamma t}$& $4 \gamma$\\
\hline
pbc, odd $L$ &$0$& $2$ & 0 & $\frac{1}{L\sqrt{8 \pi t}}$ &$\frac{1}{2}\ln ( t L^2)+\frac{1}{2}\ln (8 \pi)$\\
\hline
obc, any $L$ &$0$& $2$ & 0 & $\frac{1}{L\sqrt{8 \pi t}}$ &$\frac{1}{2}\ln ( t L^2)+\frac{1}{2}\ln (8 \pi)$\\
\hline
obc, any $L$ &$W$& $2$ & 0 & $\frac{\sqrt{\pi}}{2 L\sqrt{a t}}$&$\frac{1}{2}\ln (tL^2)+\frac{1}{2}\ln \frac{4a}{\pi}$\\
\hline
pbc, any $L$ &$W$& $2$ & 2 & $\frac{\sqrt{\pi}}{2 (at)^{3/2}}$ &$\frac{3}{2}\ln t+\frac{1}{2}\ln \frac{4 a^3}{\pi}$\\
\hline
\end{tabular}
\caption{Decay of $I(t)$ in regime IV for different cases; $\alpha$ and $\beta$ are exponents defined in Equation~\eqref{alphabeta}. For obc, the scaling of $I(t)$ with $t$ does not qualitatively change on addition of disorder, whereas for pbc they are remarkably different. See text for details.} 
\label{table1}
\end{table*}
In this section, we shall discuss the behavior of $I(t)$ in the fourth time window, where we observe an asymptotic power law decay irrespective of the nature of disorder distribution. This is the first regime where we see $L$ dependent behavior, as evident from Fig.~\ref{fig1}(b), where the evolution of $-\ln I(t)$ is plotted for different $L$'s with open boundary conditions. Additionally, unlike the previous regimes which did not depend on the boundary conditions, behavior of $I(t)$ in region IV is strongly dependent on such factors. As shown in Fig.~\ref{fig4}(c), for open boundary conditions, we see a $I(t)\propto 1/(L\sqrt{t})$ or $-\ln I(t)\sim \frac{1}{2}\ln (tL^2)$ behavior(dependence on $\gamma$ is more complicated and discussed later in the section). However for periodic boundary conditions (pbc), as plotted in Fig.~\ref{fig4}(d), $-\ln I(t)\sim \frac{3}{2}\ln t$, show a completely different behavior with a different exponent and no system size dependence.

 To explain this behavior, we first realize that in this regime charge is transported on longer scales and the system can no longer be described by three-site models of the previous section. Hence we need to study the eigenspectrum of the Liouvillian. Additionally, we need to focus on the low magnitude eigenvalues as this regime is asymptotic. However, since we have the hierarchical structure of observables, we do not need to study the full $4^L \times 4^L$ Liouvillian, but the eigenspectrum of $\mathcal{Q}$ in Equation~\eqref{linearised}, which, as mentioned earlier, is related to the one-particle Liouvillian of the problem. We can write the solution of Equation~\eqref{linearised} generically as,
\begin{equation}
\mathbf{f}(t)=\mathbf{f}_{{\rm NESS}}+\sum_j \mathbf{\delta f}(j) e^{(\lambda_j+i \Omega_j) t},
\label{rhoness}
\end{equation}
where $\mathbf{f}$ is defined under Equation~\eqref{linearised}, $\mathbf{f}_{{\rm NESS}}$ is the steady state value of the observables, $\lambda_j$ and $\Omega_j$ are the real and imaginary parts of the eigenvalues of $\mathcal{Q}$ and $\mathbf{\delta f}(j)$ contains the corresponding eigenvectors weighted by the initial conditions. Typically, the weights are similar to the corresponding elements of the eigenvector, i.e. $\mathbf{\delta f}(j) \sim m_j^2$ where $m_j$ is the $j^{{\rm th}}$ eigenvector (see Appendix~\ref{appC} for more details).

From Equation~\eqref{rhoness}, it might naively seem that any observable should show exponential decay with different rates at different timescales, depending on $\lambda_j$'s. However, in reality, in the thermodynamic limit the eigenspectrum becomes continuous, and this leads to a power law approach of observables towards NESS. To demonstrate this, assume without loss of generality that
\begin{equation}
 \lambda_j\sim -j^\alpha, \hspace{0.2in} \mathbf{\delta f}(j) \sim j^{\beta}
 \label{alphabeta}
\end{equation} and\footnote{$\Omega_j$ gives an oscillatory term which is usually $\sim 0$ for the relevant low magnitude eigenvalues of the Liouvillian spectrum. If they are included the power law reduces by 1\cite{MedvKehrinPhysRevB.90.205410}} $\Omega_j \ll \lambda_j$. Then Equation~\eqref{rhoness} can be written in the continuum limit as~\cite{CaiPhysRevLett.124.130602,CaiBarthelPhysRevLett.111.150403,MedvKehrinPhysRevB.90.205410}
\begin{equation}
\mathbf{f}(t)-\mathbf{f}_{{\rm NESS}}\sim\int dj j^{\beta} e^{-j^\alpha t}\sim t^{-\frac{\beta+1}{\alpha}}.
\label{scaling}
\end{equation}
 In what follows, we shall understand the behavior of $I(t)$ for different systems using Equation~\eqref{rhoness}. 

In Table~\ref{table1} we first summarize the results of various cases which shall be discussed in this section.

Since $\mathbf{f}$ contains all the two particle fermionic operators in the system, while $I(t)=\frac{1}{L}\sum_{k=1}^L(-1)^{k+1} C_{kk}(t)$, we can rewrite Equation~\eqref{rhoness} in the continuum for $I(t)$ as
\begin{equation}
I(t)=I_{{\rm NESS}}+\int dj \mathcal{I}(j) e^{\lambda_j t}
\end{equation}
where  
\begin{equation}
\mathcal{I}(j)=\frac{1}{L}\sum_{k=1}^L (-1)^{k+1} [\delta \mathbf{f}(j)]_{(k-1)L+k},
\label{support}
\end{equation} sums over the relevant operator subspace for the finite $L$ system. Equation~\eqref{scaling} gets appropriately modified.

Let us first look at the simpler case of systems without disorder. It is known that the low lying eigenenergies in the one-particle Liouvillian, in open boundary conditions are approximately \cite{MarkoPhysRevE.92.042143} 
$\lambda_j=2 \left(\sqrt{4 \cos \left(\frac{ \pi  j}{L}\right)-3}-1\right), j=0\hdots L-1$. 
For periodic boundary conditions the expression becomes, $2 \left(\sqrt{4 \cos \left(\frac{ 2 \pi  j}{L}\right)-3}-1\right), j=0\hdots L-1$, i.e. there is a degeneracy of $2$. Since we are interested in the regime $j/L \ll 1$, we can approximate  obc eigenvalues as $\frac{2 \pi^2 j^2}{L^2 \gamma}$. For pbc this gets multiplied by a factor of $4$ and hence in both cases $\alpha=2$.

$\mathcal{I}(j)$ can be computed from the relevant elements of the eigenvectors of the one-particle Liouvillian. It turns out that the relevant elements of the $j^{{\rm th}}$ eigenvector can be approximated to be exactly the free fermion wavefunction with the corresponding boundary condition, i.e. for pbc, they are  $\frac{1}{\sqrt{L}}\sum_{k=1}^{L}\exp(2 i \pi j k/L)$. Hence $\mathcal{I}(j)\sim\frac{1}{L}[\sum_{k=1}^{L}\frac{1}{\sqrt{L}} \cos(\pi k)\exp(2 i \pi j k/L)]^2$ is identically $0$ for even $L$ due to the symmetry of the wavefunction.\footnote{The square comes from the weightage of the initial conditions, whose weights are of the same order as the element of the eigenfunction. Hence we use $\sim$ instead of $=$.} Since $I$ has no support on the low magnitude eigenspectrum of the Liouvillian, it does not show a power law decay. Instead it decays exponentially with the largest $\lambda_j=4 \gamma$.

The situation for odd $L$ in pbc is different as the one site is unpaired. In this case $\mathcal{I}(j)\sim\frac{1}{L}[\sum_{k=1}^{L} \frac{1}{\sqrt{L}} \cos(\pi k)\exp(2 i \pi j k/L)]^2 \sim \frac{1}{L^2}$ for $j \ll L$. This means $\beta=0$, and hence $I(t)\sim 2\int dj \frac{1}{L^2} e^{-8 \pi^2 j^2 /L^2}= \frac{1}{L\sqrt{8 \pi t}}$, where the factor $2$ in front comes from the degeneracy of the eigenvalues. 

For obc the eigenvectors are given by, $\sqrt{\frac{2}{L}}\sum_{k=1}^{L}\cos( \pi j k/L)$. Hence $\mathcal{I}(j)=\frac{1}{L}[\sum_{k=1}^{L}\sqrt{\frac{2}{L}} \cos(\pi k)\cos( \pi j k/L)]^2$. This equals  $0$ when both $L$ and $j$ are even or odd, and $\frac{2}{L^2}$ otherwise. This gives $I(t)\sim \int dj \frac{2}{ L^2} e^{-2 \pi^2 (2j)^2 /L^2}= \frac{1}{L\sqrt{8 \pi t}}$. These results for the clean systems have been matched with exact numerics, data not shown.

Now we focus our attention to disordered systems. Intuitively, we can predict that since the behavior in this regime involves the low energy eigenspectrum of the Liouvillian, it should not drastically change on addition of disorder. If we look into the obc case, we indeed see $ \propto 1/(L\sqrt{t})$ behavior for both clean and disordered systems, but not so for the pbc case where we see a $t^{-3/2}$ behavior.
\begin{figure}
\includegraphics[width=0.9 \columnwidth]{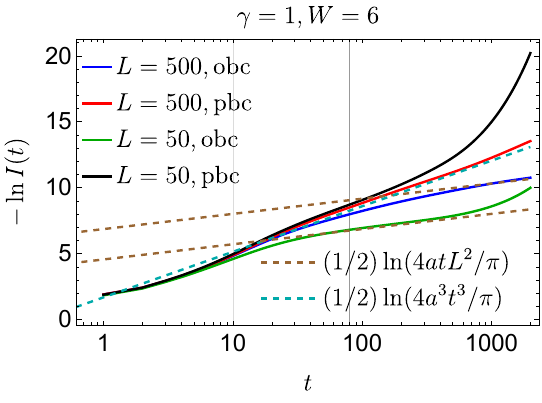}
\caption{ Comparing the behavior of $-\ln I(t)$ for different boundary conditions at different system sizes $L$. The black vertical lines approximately denote the time when obc data deviates significantly from pbc. The two brown dashed lines denote the theoretical results for obc at shown $L$, and the green dashed line denotes the pbc result.}
\label{figinsert}
\end{figure}

 Unfortunately it is very difficult to find an analytical expression for the eigenspectrum when we add the disorder in the system, so we have to resort to exact diagonalization. However, we can make some crude approximations about the low lying eigenvalues and their scaling with $L,W$ and $\gamma$ since the evolution is described by the effective Heisenberg Hamiltonian, Equation~\eqref{effham} in this time regime. First, since the effective Hamiltonian is a Heisenberg model, we conclude that the low energy distribution would be $\propto 1/L^2$. We make a further educated guess by borrowing the result $\lambda_j=\frac{2 \pi^2 j^2}{L^2 \gamma}$ for clean systems from Ref.~\onlinecite{MarkoPhysRevE.92.042143} and then introducing the relevant prefactor in $\mathcal{H}_{{\rm eff}}$ for disordered systems. Thus, we have disorder averaged
 \begin{equation}
 \lambda_j \sim -a(\gamma,W) j^2/L^2,
 \label{eigenvaluesapp}
\end{equation}
for $j \ll L$, where
 \begin{equation}
 a(\gamma, W)=\frac{8 \pi^2 \gamma }{ (\frac{2W^2}{3}+4 \gamma^2)}
 \label{a}
\end{equation}
and we have used $\langle (h_{j}-h_{j+1})^2 \rangle\sim 2 W^2/3$ for the box distribution $(-W,W)$.  Henceforth we shall write $a(\gamma, W)$ as $a$.
\begin{figure}
\includegraphics[width=0.9 \columnwidth]{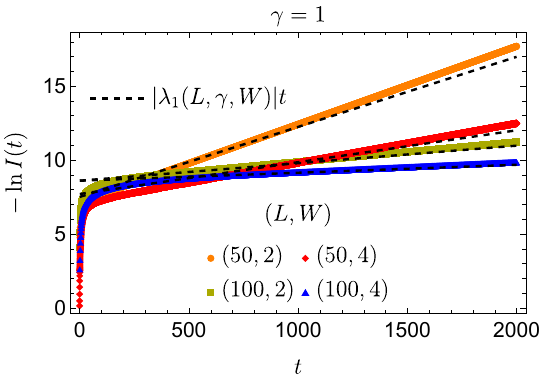}
\caption{Plot showing the linear growth of $-\ln I(t)$ in region V for obc finite size systems of different lengths $L$ and at disorder strengths $W$. The black dashed lines denote the expected growth of $-\ln I(t)$ with the rate given by corresponding $\lambda_1$ dependent on $L,W$ and $\gamma$.}
\label{fig5a}
\end{figure}
 \begin{figure*}
\includegraphics[width=0.9 \columnwidth]{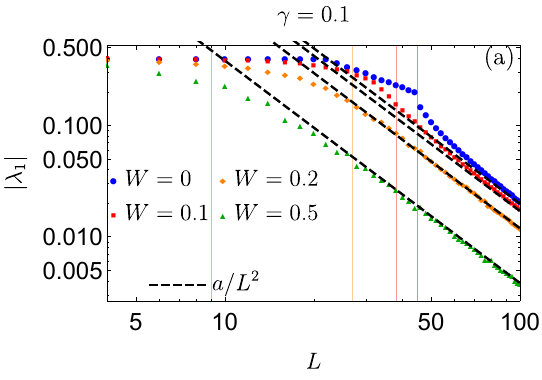}
\includegraphics[width=0.9 \columnwidth]{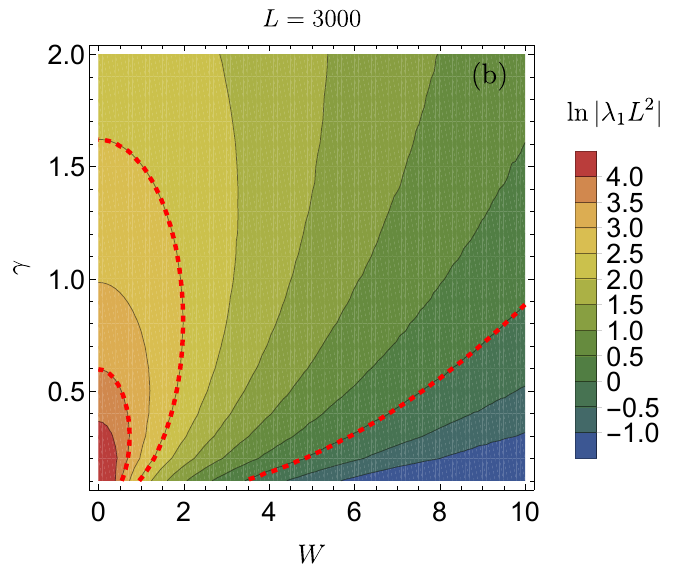}

\caption{ Scaling of $\lambda_1$, the smallest Liouvillian gap, with different parameters for obc. (a) Plot showing $1/L^2$ scaling of $\lambda_1$, for different disorder strengths, $W$. $L_c$ from Equation~\eqref{criticallength} is marked by appropriate colored gridlines in the plot. Averaging is done over $100$ realizations.  (b) Plot showing dependence of $\ln |\lambda_1 L^2|$ with $W$ and $\gamma$ for $L=3000$ averaged over 100 disorder realizations. The red dashed lines indicate three constant contours at $0,2.5$ and $3.5$ for $\ln a$ obtained from Equation~\eqref{a}.}
\label{fig5}
\end{figure*}
In Fig.~\ref{fig4}(a) we see a very good agreement between our approximate prediction in Equation~\eqref{eigenvaluesapp} and numerical results for the low magnitude eigenvalues of the Liouvillian. Now we turn our attention to the eigenvectors. As shown before, imbalance has a constant support which was significant for odd/even $j$ in clean obc systems. Small perturbations around this value due to disorder do not affect the overall scheme of things upon averaging over many disorder realizations, as can be verified from Fig.~\ref{fig4}(b). As seen in Fig.~\ref{fig4}(b), for odd $j$, $\mathcal{I}_j$ is almost the same for clean and disordered obc systems with even $L$. For even $j$ on the other hand this quantity is no longer identically $0$, as the symmetry of the system is broken due to disorder and shows a $(\sqrt 2j)^2/L^2$ scaling. However, this does not seriously affect the evolution as their magnitude is much smaller than for odd $j$. Consequently, for disordered obc systems,
\begin{eqnarray}
I(t)\sim&& \frac{1}{L} \int dj \frac{2}{L} e^{-a (2j)^2 /L^2}\nonumber \\
&&= \frac{\sqrt{\pi}}{2L\sqrt{a t}}.
\label{obclaw}
\end{eqnarray} 
This is the behavior seen in Fig.~\ref{fig4}(c), where disorder, $W=4$ , is chosen to be smaller to highlight region IV (data for $W=10$ is similar at larger times and not shown). 

  However, for the periodic case, we have a different scenario. Previously $\mathcal{I}(j)$ had no support for any $j$ in this regime for clean systems, but now, due to breaking of symmetry, $\mathcal{I}(j)\sim (\sqrt{2}j)^2/L^2$ as evident from Fig.~\ref{fig4}(b). The eigenvalues as seen in Fig.~\ref{fig4}(a) follow the same scaling law as obc (the degeneracy in the smallest magnitude eigenvalue is resolved as we increase $j$). Hence we get
  \begin{eqnarray}
  I(t)\sim&& \frac{1}{L} \int dj \frac{2j^2}{L^2} e^{-a (j)^2 /L^2}\nonumber \\
  =&& \frac{\sqrt{\pi}}{2 (a t)^{3/2}}.
  \label{pbclaw}
   \end{eqnarray} Indeed this is what we see in Fig.~\ref{fig4}(d), including the lack of $L$ dependence in the result. For smaller $W$, e.g. $W=4$, using theoretical $a$ in Equation~(\ref{a}) differs slightly from numerical data (not shown); however, time dependence is still $1/t^{3/2}$.

We can see that the difference between open and periodic boundary conditions comes due to different scaling of the overlaps of imbalance with eigenvectors. On a more physical note it is worth observing that the imbalance dynamics will agree between open and boundary conditions at least up to extensive times $ t \propto L$ when the causal cone hits a boundary and the system starts to feel the boundary condition. This can be seen in Fig.~\ref{figinsert}, where we compare dynamics for two types of boundary conditions. We can see that indeed for $L=50$ the obc and pbc data overlap until around $t\sim 10$, while for $L=500$ there is overlap until almost an order of magnitude larger, $t \sim 80$.

\subsection{Region V}
\label{region5}
As mentioned in Sec.~\ref{summary}, region IV becomes asymptotic in the thermodynamic limit, when we have a continuum in the eigenspectrum. For finite size systems though, the spectrum eventually gets resolved and hence for larger $t\sim L^2W^2/\gamma$ we observe a different behavior, which we classify as region V. In this region, $I(t)$ shows an exponential decay, with the rate governed by the Liouvillian gap $\lambda_1$.\footnote{the exception being clean pbc systems with even $L$ where imbalance has no support on the eigenfunction as shown before} $\lambda_1$ is the largest real part of the non zero Liouvillian eigenvalue for both obc and pbc systems. Clean obc, odd L systems constitute an exception, as the determination of the imbalance operator is based on the next largest eigenvalue, since the largest non zero eigenvalue has zero support. In Fig.~\ref{fig5a}, we see that $-\ln I(t)$ grows linearly with time at large times for disordered obc systems, with the slope $|\lambda_1(L,\gamma,W)|$ [which is approximated by Equation~\eqref{a} with $j=1$, explained in the following], verifying our expectation.  The plots are similar for pbc (data not shown), with slope approximately four times corresponding obc systems as can be seen by comparing $\lambda_1$ for pbc and obc in Fig.~\ref{fig4}(a). As before, analytical results for clean systems are well known~\cite{MarkoPhysRevE.92.042143}, but upon addition of disorder one can only resort to numerics for exact results. Nevertheless, the description given by Equation~\eqref{effham} is still valid in this time scale, and explains the scaling behavior of $\lambda_1$ with $L,W$ and $\gamma$ as discussed below.

In Fig.~\ref{fig5}(a) we show the change of $\lambda_1$ with $L$ for constant $W$ and $\gamma$. It is known that the $1/L^2$ scaling of $\lambda_1$ works for sufficiently large $L$ for clean systems. In fact, it has already been shown in Ref.~\onlinecite{MarkoPhysRevE.92.042143} that the critical length required to observe this scaling is $L_c \sim \frac{\pi \sqrt{2}}{\gamma}$. For $\gamma=0.1$ this is at $L\sim 45$ and is shown by the blue gridline in Fig.~\ref{fig5}(a). This behavior also continues when we introduce disorder in the system. It is evident from the figure that upon increasing $W$, $L_c$ decreases. This can be explained as in the previous section using Eq~\ref{effham} and Equation~\eqref{eigenvaluesapp}, which tells us that, on addition of disorder of strength $W$, $\gamma$ can be effectively replaced by $(4\gamma^2+\frac{2 W^2}{3})/(4\gamma)$. This  gives
\begin{equation}
L_c(W)\sim \frac{ 4 \sqrt{2} \pi \gamma}{4\gamma^2+\frac{2 W^2}{3}},
\label{criticallength}
\end{equation}  shown as different colored gridlines in the plot, and capture the transition point between the two regimes quite well.

Finally in Fig.~\ref{fig5}(b) we show the change of $\lambda_1$ on changing $W$ and $\gamma$ for constant $L=3000$ via a contour plot. We expect $|\lambda_1 L^2| \sim a$ and hence the equation of constant contours can be obtained from Equation~\eqref{a} as 
\begin{equation}
\ln a={\rm constant}.
\label{fitgap}
\end{equation}
Three such contours are plotted as red dashed lines in Fig.~\ref{fig5}(b) and they match very well with the contours obtained from exact numerics.

\section{Discussion}
\label{discussion}
In this work we have studied the evolution of imbalance, $I(t)$ for the disordered XX chain with on-site dephasing, starting from the N\'eel initial state. Using the hierarchical nature of equations for the observables, we have computed $I(t)$ for large system sizes ($L\sim 10^3$) and long times ($t \sim 10^3$). Our analysis showed the emergence of five timescales in disordered systems. 

The shortest time scale, $t < t_0 \sim {\rm Min(1/\gamma,1/W)}$ where $ I(t) \sim (1-\eta t^2)$ constitutes the  linear response regime of the system. Then, due to the localization via disorder, two timescales denoted by regions II and III emerge, which are absent in clean systems. Region II is the linear regime of what has been called a stretched exponential decay in previous works\cite{AltmanPhysRevLett.116.160401,CaiBarthelPhysRevLett.111.150403,CaiPhysRevLett.124.130602,SchneiderPhysRevB.106.134211,PhysRevLett.116.237203Garrahan}, where $I(t) \sim (1-\beta \sqrt{t})$, and is universal irrespective of the nature of disorder chosen, and continues until $t_1 \sim 0.1 W^2/8\gamma$, before smoothly transitioning to a disorder dependent behavior in region III which continues until $t_2 \sim 5 W^2/(8\gamma J^2)$. We thus conclude that the stretched exponential fits are neither universal nor asymptotic for the system under study. Simple local three-site models describe the behavior in these two regimes and a rescaled time emerges during the analysis $\tau= 8 \gamma t/W^2$. An effective Heisenberg model given by Equation~\eqref{effham} also provides the correct description of the evolution from region II. 

For $t > t_2$, where region IV begins, dephasing breaks localization and the system shows size and boundary condition dependent power law decay of $I(t)$. In fact for pbc it shows a $1/\sqrt{t^3}$ decay, while for obc the initial decay $1/\sqrt{t^3}$ transitions at time $t \propto L$ into $1/(L\sqrt{t})$. These were explained from the continuum limit of the low magnitude eigenspectrum of the Liouvillian. Finally, we discussed that for finite sized systems due to resolution of the eigenspectrum, the decay of $I(t)$ in the final region V at $t>t_3$ is exponential with the rate governed by the relevant Liouvillian gap $\lambda_1$. We also provided analysis of how $\lambda_1$ typically scales with $W,L,\gamma$ using the effective Heisenberg Hamiltonian approach.

Our analysis gives insight to the mechanism behind the evolution of imbalance in such a system. It demystifies the behavior of the system in regions II and III clearly showing its local origin and gives us the ability to make predictions about the evolution for many different disorder distributions. Additionally, the proper analysis of the Liouvillian eigenspectrum provides us with the correct asymptotic description of evolution for different models and boundary conditions. 

We would also like to briefly mention the nuances of crossover from one regime to another. Since the transition is smooth, there is a always finite time taken by the system to go from one regime to another. For example, from region III to IV, where the finite size effects first appear, the system typically takes $t\sim L$ to transition to the asymptotic power law for open boundary conditions. However, this does not seem to be the case for periodic boundary conditions.

 Additionally, it is worthy to remark that the effective Hamiltonian description~\cite{MariyaMarkoPhysRevB.93.094205} of Equation~\eqref{effham} is also valid for weakly interacting systems where the interaction strength is much smaller than disorder strength. This means weak interaction does not play much of a role in evolution in such timescales, so we expect most of our results to hold for weakly interacting systems as well.
 
Furthermore, while we have discussed the N\'eel initial state in this work, decay of $I(t)$ from other generic computational initial states also share some similar features. The behavior in regions I, IV and V seen for the N\'eel state is still observed for other typical states. The differences occur in $t_2$ and the behavior in regions II and III. The first difference in other initial states is that $I(t=0)< 1$ i.e. $-\ln I(t=0)>0$. Since $t_2$ is defined by a finite non zero value of $I$ universal for a given disorder distribution, relaxation from different $I(t=0)$ typically reaches this value at different times. Secondly, all the sites of the initial state are not locally equivalent unlike the N\'eel state. Hence, the scaling in regions II and III show a difference due to variation in local behaviors. A detailed analysis of this aspect is beyond the scope of the present work. 

There are still a few open questions left in the context of this work. One natural extension would be to study the evolution of other observables such as current in such a model. The question of whether the scaling laws depend on the type of dephasing is also worth studying. Indeed, the hierarchical structure of observables or the effective Heisenberg model might break down if we choose a different model of dephasing. Additionally, while regions I--III have been studied under different disorder settings, we have provided a general explanation for the power law decay in region IV, discussing the two cases where it shows different exponents. Specific potentials may give rise to unique behavior in this time scale, and an analysis of that is left for a future work.

\section*{acknowledgement}
We would like to acknowledge support by Grants No.~J1-1698, No.~J1-4385 and
No.~P1-0402 from the Slovenian Research Agency. RG would also like to acknowledge support from UKRI Grant No.~EP/R029075/1. Comments by an anonymous referee that helped us improve the presentation are gratefully acknowledged.

\bibliography{xxdephasing} 
\bibliographystyle{apsrev4-2}
\appendix
 
\section{Two site model}  
 \label{appA1}
In this appendix we shall derive the eigenspectrum expressions for the two-site model using Equation~\eqref{linearised} as the starting point of our computation. For regime I this simple model is enough to qualitatively explain the features seen.
For the two site model, $\mathbf{f}=(C_{11},C_{12},C_{21},C_{22})$ and
\begin{equation}
\mathcal{Q}=\left(
\begin{array}{cccc}
 0 & 2 i   & -2 i  & 0 \\
 2 i   & 2 i \delta+4 \gamma  & 0 & -2 i   \\
 -2 i  & 0 & -2 i \delta+4 \gamma  & 2 i  \\
 0 & -2 i  & 2 i   & 0 \\
\end{array}
\right)\label{matrix}
\end{equation} 
where we have taken $J=1$ and $\delta=h_1-h_2$. For $\delta=0$ (clean system), one can compute the eigenvalues exactly as ,
\begin{equation}
\left[0,4 \gamma ,2 \left(\gamma -\sqrt{\gamma ^2-4}\right),2 \left(\sqrt{\gamma ^2-4}+\gamma \right)\right],
\end{equation} and the time dependent solutions to the $C_{11(22)}$ can be written as,
\begin{eqnarray}
C_{11(22)}&=&e^{2 \gamma  t} [(-)\cosh \left(2 t \sqrt{\gamma ^2-4 }\right)\nonumber \\
&&-(+)\frac{\gamma  \sinh \left(2 t \sqrt{\gamma ^2-4 }\right)}{\sqrt{\gamma ^2-4}}].
\end{eqnarray}

Hence,
\begin{equation}
I(t)=\frac{1}{2}(C_{11}(t)-C_{22}(t))\sim 1- 8 t^2 +O(t^3).
\end{equation}
This is the demonstration of the short time quadratic behavior when $t<1/\sqrt{\gamma^2-4}$. 

However when $\delta>0$ the exact solution is given from a cubic equation since one of the eigenvalues of $\mathcal{Q}$ is always $0$. The expressions are complicated and hence not shown. But we can formulate a simple perturbative result for $\delta \gg1,\gamma$. To do so we first rearrange $\mathcal{Q}$ to separate the degenerate block and non-degenerate blocks as
\begin{equation}
\mathcal{Q}=\left(
\begin{array}{cccc}
 0 & 0 & 2 i   & -2 i   \\
 0 & 0 & -2 i   & 2 i \ \\
 2 i  & -2 i  & 4 \gamma +2 i \delta  & 0 \\
 -2 i   & 2 i  & 0 & 4 \gamma -2 i \delta  \\
\end{array}
\right)=\begin{pmatrix}
\bf{O}_{2 \times 2} & \bf{B}_{2 \times 2} \\
\bf{C}_{2 \times 2} &\bf{D}_{2 \times 2}
\end{pmatrix} .
\label{rearrange}
\end{equation}
Then applying second order degenerate perturbation theory to $\bf{O}$ and second order non degenerate perturbation to $\bf{D}$ we have the eigenvalues as,
\begin{equation}
\left[0,\frac{16 \gamma }{4 \gamma ^2+\delta ^2},4 \gamma +2 i \delta +\frac{8}{4 \gamma +2 i \delta },4 \gamma- 2 i \delta-\frac{8}{-4 \gamma +2 i \delta }\right]
\end{equation}
and the solution for $\delta \gg \gamma$ is given by
\begin{equation}
I(t)=-\frac{4 \cos \left(2 \sqrt{\delta^2+4} t\right)+\delta^2}{\delta^2+4}=1-8 t^2+O\left(t^4\right)
\end{equation}
where the quadratic decay with $t$ remains but it is valid until $t \sim 1/\sqrt{\delta^2+4}$.

\section{Three site model}
\label{appA2} 
\begin{figure}
\includegraphics[width=0.9 \columnwidth]{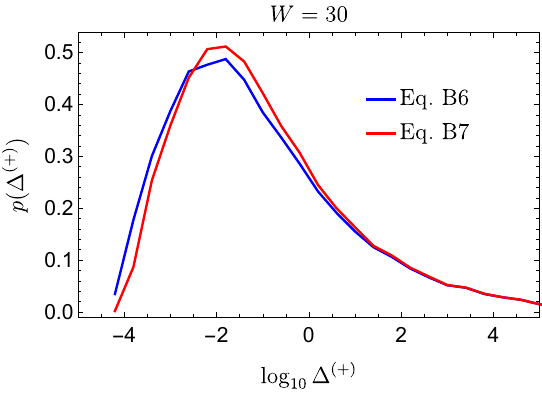}
\caption{Comparison of distribution of $p(\Delta^{(+)})$, between $\Delta^{(+)}$ given by Equation~\eqref{appeq1} and approximated by Equation~\eqref{approx}.}
\label{fig6a}
\end{figure}
Now we repeat the above calculation for a three site model which describes the behavior in regimes II and III. 
In this case $\mathbf{f}$ is given by Equation~\eqref{threesitef}, while $\mathcal{Q}$ is given by
\begin{equation}
\left(
\begin{array}{ccccccc}
 0 & 2 i   & -2 i   & 0 & 0 & 0 & 0 \\
 2 i  & 4 \gamma +2 i \delta_1 & 0 & -2 i   & 0 & 0 & 0 \\
 -2 i   & 0 & 4 \gamma -2 i \delta_1 & 2 i   & 0 & 0 & 0 \\
 0 & -2 i  & 2 i  & 0 & 2 i \tau  & -2 i   & 0 \\
 0 & 0 & 0 & 2 i   & 4 \gamma +2 i \delta_2 & 0 & -2 i   \\
 0 & 0 & 0 & -2 i  & 0 & 4 \gamma -2 i \delta_2 & 2 i  \\
 0 & 0 & 0 & 0 & -2 i   & 2 i  & 0 \\
\end{array}
\right)
\end{equation}
where  $h_1-h_2=\delta_1$ and $h_2-h_3=\delta_2$.  Rearranging in the form of Equation~\eqref{rearrange}, we get the general form in this case as
\begin{equation}
\mathcal{Q}=\begin{pmatrix}
\bf{O}_{3 \times 3} & \bf{B}_{3 \times 4} \\
\bf{C}_{4 \times 3} &\bf{D}_{4 \times 4}
\end{pmatrix} .
\end{equation}
Then using second order degenerate perturbation theory in the subspace of $\bf{O}$, we can compute the eigenvalues as $\Delta^{(0)}=0$ and 
 \begin{widetext}
 \begin{equation}
\Delta^{\pm}=\frac{8 \gamma  \left(8 \gamma ^2\pm\sqrt{16 \gamma ^4+4 \gamma ^2 \left(\delta_1^2+\delta_2^2\right)+\delta_1^4-\delta_1^2\delta_2^2+\delta_2^4}+\delta_1^2+\delta_2^2\right)}{\left(4 \gamma ^2+\delta_1^2\right) \left(4 \gamma ^2+\delta_2^2\right)}.
\label{eigenvalues}
\end{equation} 
\end{widetext}
Since we are in the regime where $W \gg \gamma$, we have $\delta_1$, $\delta_2 \gg \gamma$, and hence we can approximate Equation~\eqref{eigenvalues} as Equation~\eqref{approxeigen}.

One can also compute the eigenvectors in this subspace and plug in the initial N\'eel state to find the coefficients,
\begin{eqnarray}
d^{(0)}&=& -\frac{1}{3}\nonumber \\
d^{(+)}&=&\frac{(2 \delta_1^2-\delta_2^2+\kappa)(\delta_2^2+\kappa)}{3 \delta_1^2 \kappa}\nonumber \\
d^{(-)}&=&\frac{(2 \delta_1^2-\delta_2^2-\kappa)(-\delta_2^2+\kappa)}{3 \delta_1^2 \kappa}
\end{eqnarray}
where $\kappa=\sqrt{\delta_1^4-\delta_1^2 \delta_2^2+\delta_2^4}$. We can see that when $|\delta_1|=|\delta_2|$, $d^{(+)}=\frac{4}{3},d^{(-)}=0$. In the opposite regime, i.e. when $|\delta_1^2 - \delta_2^2|\gg 0$, $d^{(+)} \rightarrow 1$ and $d^{(-)}\rightarrow -d^{(0)}$. Consequently, it can be shown that $1 \le d^{(+)} \le 4/3$, $0 \le d^{(-)} \le 1/3$ and hence $d^{(+)}$ always provides the largest contribution to the evolution.

We use the result for $|\delta_1|=|\delta_2|$ when we compute $I(t)$ for alternating potential plotted in Fig.~\ref{fig3}(b) as
\begin{eqnarray}
I(t)&=&-\frac{1}{3}+\frac{4}{3} e^{-\Delta^{(+)} t}\nonumber \\
&\sim &e^{-4/3 \Delta ^{(+)} t}=e^{-\tau},
\label{alternating2}
\end{eqnarray}
where we have used $\Delta ^{(+)}\sim 6 \gamma /W^2$ for $W \gg \gamma$. One surprising aspect is that Equation~\eqref{alternating2} is a good fit to exact numerics in Fig.~\ref{fig3}(b) beyond $\tau \sim 1$. This can be explained by observing that if we take local models of higher sizes, then we will find numerically that $d^{(0)}$ decreases and the weight gets shifted to a mode which evolves with $\sim4/3 \Delta ^{(+)}$. Thus Equation~\eqref{alternating2} remains valid even when the transport occurs in a more non local region (but not in the full lattice) and hence to longer times, before the system shows the asymptotic power law behavior. Also notice that because there is only one mode of relaxation, the decay is exponential in this timescale.

For the random disorder case we will have a distribution of $d$ and $\Delta^{(\pm)}$ based on the distribution of disorder. However, the principal mode of decay is via $\Delta^{(+)}$. To the leading order, the effect of $d_0$ on this mode becomes progressively smaller as we move away from the limiting alternating potential case $|\delta_1|=|\delta_2|$, as it is countered by the $d^{(-)}$ term. Furthermore, $\Delta^{(-)}$ is typically too small to have a significant effect on evolution in regimes II and III. Finally, the shifting of weight towards a mode $\sim \Delta^{(+)}$ when we take slightly bigger but still local models is valid for the disordered case as well. Hence simply averaging over $\Delta^{(+)}$ modes is usually enough to get accurate results as presented in Fig.~\ref{fig3}(a).

The final point we need to address is the validity of the approximation we have used to arrive at Equation~\eqref{master1} from Equation~\eqref{firstapprox}. Let us recall that
\begin{equation}
\Delta^{(+)}\sim\frac{8 \gamma  (\delta_1^2+\delta_2^2+\sqrt{\delta_1^4+\delta_2^4-\delta_1^2\delta_2^2})}{\delta_1^2 \delta_2^2}.
\label{appeq1}
\end{equation}
For very large $W$, we can expect terms with $|\delta_1|\sim |\delta_2|$ would be statistically insignificant. Hence to get a tractable expression, we take the approximation $|\delta_1^2 - \delta_2^2|\gg 0$. This would constitute the long tails of the distribution of $\Delta^{(+)}$. Finally, to be able to perform the analytical computation simply, we choose to approximate $\Delta^{(+)}$ by
\begin{eqnarray}
\Delta^{(+)}&\sim &\frac{16 \gamma(\delta_1^2+\delta_2^2)}{\delta_1^2 \delta_2^2}
\label{approx}
\end{eqnarray}
 and show its agreement with full $\Delta^{(+)}$  in Fig.~\ref{fig6a}.
The approximation qualitatively represents the exact curve reasonably well. Also, since it becomes exact in the tail, i.e.. for large $\Delta^{(+)}$, consequently the expression computed using these results represents the exact result for $I(t)$ for smaller $t$ more accurately than the rest, a feature seen in Fig.~\ref{fig3}(a) and Fig.~\ref{fig3}(b). The exact stretched exponential result which arises out of this approximation for Gaussian disorders is also valid only for small $t$.
 \section{Agreement of exact numerics with evolution by $\mathcal{H}_{\rm eff}$}
 \label{appB}
 \begin{figure}
\includegraphics[width=0.9 \columnwidth]{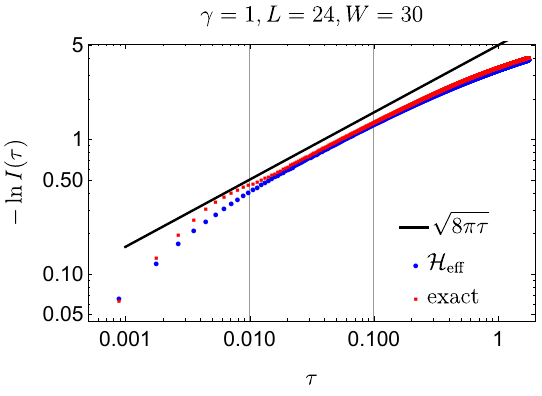}
\caption{ Comparison between exact numerical results for evolution of $I(t)$ with those obtained by solving Equation~\eqref{Heffevolution} averaged over 20 realizations for $L=24$ and $W=30$, obc.}
\label{fig6}
\end{figure}
In the end of Sec.~\ref{region23} we mentioned that we could use an effective Hamiltonian to replicate the behavior of $I(t)$ from a bit later than $\tau=\tau_0$. Let us discuss this aspect in a bit of detail.

Following Ref.~\onlinecite{MariyaMarkoPhysRevB.93.094205}, at a sufficiently large timescale the density matrix becomes effectively diagonal and the evolution of the diagonal elements $\rho_{i}$ is given by,
\begin{equation}
\frac{d\rho_{i}(t)}{dt}=\sum_j[\mathcal{H}_{{\rm eff}}]_{ij}\rho_{j}(t)
\label{Heffevolution}
\end{equation}
where $\mathcal{H}_{{\rm eff}}$ is given in  Equation~\eqref{effham}. $I(t)$ can then be computed from $\rho_i(t)$ as 
\begin{equation}
I(t)=\sum_i^{2^L}\rho_i(t) \sum_j^L\bra{\alpha_i}(-1)^j \sigma_j^z\ket{\alpha_i}
\label{Imbalanceheff}
\end{equation} 
where $\ket{\alpha_i}$ can be taken as the computational basis states. We plot the comparison of results obtained from exact numerics using Equation~\eqref{master1} and evolution of the density matrix using Equation~\eqref{Heffevolution} and Equation~\eqref{Imbalanceheff} in Fig.~\ref{fig6}. We see that from  $\tau \sim 0.01$ the two results are almost equal. This effective description reproduces the low energy eigenspectrum of the Liouvillian quite accurately and hence correctly approximates the evolution from $\tau \gg \tau_0$, and thus is effective in capturing the evolution of $I(t)$ from approximately region II, as expected.
 \section{Computation of low magnitude eigenspectrum of Liouvillian}
 \label{appC}
 \begin{figure}
\includegraphics[width=0.9 \columnwidth]{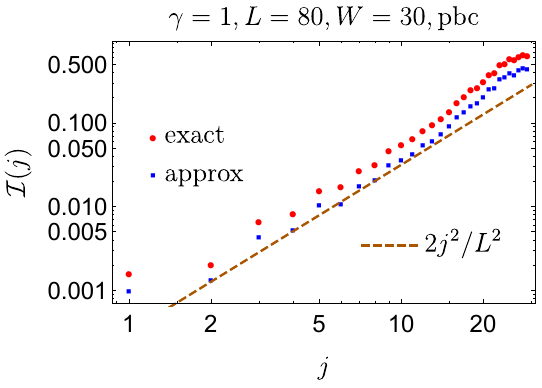}
\caption{Comparison of exactly computed $\mathcal{I}_j$ via solving $L\times L$ linear equations for the initial conditions with our approximation using the squares of the eigenvector elements for a system with $L=80$, $W=30$, $\gamma=1$ and periodic boundary conditions, averaged over $10^3$ disorder realizations. The dashed line is the approximate scaling we have used in Sec.~\ref{region4-1}. }
\label{fig7}
\end{figure}
  Following the arguments of Ref.~\onlinecite{MarkoPhysRevE.92.042143}, we have extracted the low magnitude eigenspectrum of the system by diagonalizing a matrix of size $\sim L \times L$ instead of $\sim L^2 \times L^2$ one-particle Liouvillian. If we denote the computational basis as $|j\rangle$, then the low magnitude eigenspectrum of the one-particle Liouvillian can be approximated by eigenspectrum of the following matrix  
\begin{equation}
\begin{pmatrix}
0 & \mathbf{R}^T & 0\\
\mathbf{R} & -4\mathbb{I} \gamma & i \mathbb{I} \mathbf{X} \\
0 & -i \mathbb{I} \mathbf{X }& -4\mathbb{I} \gamma
\end{pmatrix}
\end{equation}
 written in the basis $|j\rangle\langle j|$,
$|j\rangle\langle j+1|+|j+1\rangle\langle j|$ and $|j\rangle\langle j+1|-|j+1\rangle\langle j|$. Here $\gamma$ is the dephasing, $R_{jk}=-2\rm{i}\sqrt{2}(\delta_{j,k}-\delta_{k,j+1})$ and $X_{jk}=(h_j-h_{j+1}) \delta_{jk}$, with $h_i$ being the on-site disorders. Note that unlike the clean Hamiltonian case treated in Ref.~\onlinecite{MarkoPhysRevE.92.042143}, $|j\rangle\langle j+1|+|j+1\rangle\langle j|$ are not eigenvectors of $\mathcal{L}$ as disorder breaks translational invariance.  Thus we need to effectively solve a $3L-2 \times 3L-2$ [$3L \times 3L$] problem for obc [pbc], to obtain the eigenspectrum, which is the ``tridiagonal" approximation. This allows us to compute the eigenspectrum for large systems and results are shown in Figs.~\ref{fig4}(a) and (b). \\

Furthermore, for small system sizes one can numerically show that the $\mathbf{\delta f}(j)$'s are proportional to the corresponding eigenvector elements and that approximating $\mathcal{I}(j)$ with the square of corresponding eigenvector terms is justified. In Fig.~\ref{fig7} we show the comparison of our approximation with exact $\mathcal{I}(j)$ found by numerically computing the correct weights due to the initial N\'eel state. We do exact diagonalization for  a system size of $L=80$ at large $W=30$ averaged over $10^3$ realizations under periodic boundary conditions to obtain the exact data. As one can see the qualitative nature of both the plots are the same; they vary by a factor of approximately $1.5$ (computed numerically). Hence in Fig.~\ref{fig4}(b) our approximation was able to qualitatively replicate the behavior of $I(t)$ with insignificant deviation from exact numerics.
\end{document}